\documentstyle[aps,prd]{revtex}
\input epsfig.sty 
\def\laq{\raise 0.4ex\hbox{$<$}\kern -0.8em\lower 0.62 ex\hbox{$\sim$}}
\def\gaq{\raise 0.4ex\hbox{$>$}\kern -0.7em\lower 0.62 ex\hbox{$\sim$}}

\begin{document}
\bibliographystyle{unsrt}

\title{Blue Spectra of Kalb-Ramond Axions and Fully
 Anisotropic String Cosmologies}

\author{Massimo Giovannini\footnote{Electronic address: 
m.giovannini@damtp.cam.ac.uk, giovan@cosmos2.phy.tufts.edu }}

\address{{\it DAMTP, Silver Street, CB3 9EW Cambridge, 
United Kingdom}}
\address{{\it  and}}
\address{{\it Institute of Cosmology, Department of Physics and Astronomy,}}
\address{{\it Tufts University, Medford, Massachusetts 02155, USA}}

\maketitle

\begin{abstract}
 The inhomogeneities associated with massless Kalb-Ramond axions can be 
amplified not only in  isotropic 
(four-dimensional) string cosmological models but also in the fully 
anisotropic case. If the  background geometry is isotropic, the axions 
(which are {\em not} part of the homogeneous background) develop, outside 
the horizon, growing modes leading, ultimately, to logarithmic
energy spectra which are ``red" in frequency and 
 {\em increase} at large distance scales. 
   We show that this 
conclusion can be evaded not only in the case of higher dimensional
 backgrounds with contracting internal dimensions but also in 
 the case of  string cosmological  scenarios which are {\em completely  
anisotropic in four dimensions}.
In this case the logarithmic
 energy spectra turn  out to be ``blue" in frequency and, 
consequently, {\em decreasing} at large distance scales. We elaborate
on anisotropic dilaton-driven models and we argue that, incidentally,
the background models leading to (or flat) logarithmic energy spectra for
 axionic fluctuations are likely to be isotropized by  the effect 
of string tension corrections. 
\end{abstract}
\vskip0.5pc
\noindent

\renewcommand{\theequation}{1.\arabic{equation}}
\setcounter{equation}{0}
\section{Introduction} 

In recent years a lot of effort has been devoted to the analysis of 
rather 
peculiar phenomenological implications   of string inspired cosmological 
models \cite{1}. 
One of the distinctive features of the low-energy string effective
 action is indeed the presence of the dilaton field whose sharp growth 
amplifies, via gravitational instability, not only the scalar and tensor
fluctuations of the geometry \cite{2} 
but also the (quantum mechanical) inhomogeneities
associated with other fields which are not part of the homogeneous 
background  like the gauge fields \cite{3,3b} and the
universal axion of string theory  \cite{4b,4c}, i.e. the
(four-dimensional) dual of the Kalb-Ramond antisymmetric tensor field
present in the (effective) low energy limit of the action. 

The main difference between the amplification of the metric
fluctuations and the amplification of other fields with no
homogeneous background turned out to be, a posteriori, the 
spectral amplitudes. In the case of metric fluctuations the growing
modes appearing in the corresponding evolution equations lead
typically, to  increasing (``violet'') spectra, whereas in the
case of Abelian gauge fields and axions the spectral slope can be much
milder. 

These results refer to the case of a  homogeneous background
geometry whose spatial section 
naturally decomposes into the direct product of two
(maximally symmetric) Euclidean sub-manifolds which will be  called, for
short, {\em external} and {\em internal}. 
This type of metric describes 
a situation of dimensional decoupling where the external dimensions
expand and the internal dimensions contract
\footnote{In the present investigation we
will always use the String frame, commenting, when needed, about the
Einstein frame description. For a general introduction to this
terminology see \cite{1}.}.
In previous studies concerning the phenomenological implications of
string inspired scenarios, the expanding  dimensions have been always taken to
be isotropic (i.e. described by a unique scale factor).
 Similarly,  the internal dimensions have been also taken to be isotropic
(this situation can happen if we compactify the internal dimensions on
a six-dimensional torus). This choice is not a limitation. In fact we
know that the dilaton ``vacuum'' solutions can be solved for arbitrary
(anisotropic) manifolds.

There  are, a priori, no reasons why the external and/or internal
sub-manifolds should be considered, separately, to be isotropic. 
In the context of the pre-big-bang scenario \cite{1} the
dilaton driven phase starts when the coupling constant and the
curvature are quite small but this does not necessarily implies the
isotropy of the expanding manifold. In fact, the study of the
occurrence of a dilaton-driven,
expanding, regime (in the string frame) can be connected to the problem
of the gravitational collapse of a stiff (perfect) fluid in general
relativity. Therefore, as it was recently shown \cite{5}, an anisotropic
dilaton-driven phase can be expected also in four dimensions. 

If the four-dimensional metric is {\em not} isotropic different
phenomenological consequences can be expected both for the metric
inhomogeneities and for the fluctuations of other fields (like the
Kalb-Ramond axions)  not contributing to  the homogeneous background. 
Concerning the metric perturbations, it is well
known that if the four-dimensional geometry is completely anisotropic
the scalar, vector and tensor modes of the corresponding fluctuations
are all coupled \cite{6}. More precisely the 
tensor modes can still be decoupled
from the scalar ones, provided the four-dimensional metric possess some
special symmetry \cite{7} (like the spherical symmetry, as assumed in 
\cite{5} or the cylindrical symmetry as suggested in \cite{8}). Similar
problems were addressed not in the framework of the pre-big-bang
evolution but in the context of a post-big-bang (Kasner) behavior of
the general relativistic solutions in the vicinity of a cosmological 
singularity \cite{9,9a}.

In this paper we will concentrate our attention on
the description
of mass-less string axions fields in a fully anisotropic four-dimensional
metric. 
Dilaton and graviton production in anisotropic string cosmological
models can be studied using techniques similar to the ones we are going
 to exploit.
There is a simple reason for this choice. In the study of the
quantum mechanical inhomogeneities amplified in string cosmology the
scalar and tensor modes of the metric become relevant at large
frequencies \cite{2} (i.e. small distance scales), whereas the axion
and gauge  perturbations seem to be more important at small
frequencies where they can have interesting effects on the cosmic
structures. Gauge field fluctuations can have interesting implications
in the context of the problem of large scale magnetic fields \cite{3} and
in the context of large scale anisotropies \cite{3b}.
Massless (Kalb-Ramond) axions interact only with couplings of
gravitational strength and, therefore, their presence is not
constrained by present tests of the equivalence principle. Moreover
the smallness of the coupling to the anomaly makes their conversion to
photons (in strong magnetic fields) negligible. Recently it was
pointed out that Kalb-Ramond axions, amplified from vacuum
fluctuations, can represent an interesting
constraint for pre-big-bang models since they could
be  invoked for the generation of the CMBR anisotropy in string
cosmology \cite{4b}. 

The purpose of our investigation is to understand which are the main
consequences of treating axionic perturbations in a fully anisotropic
four-dimensional metric. Our aim  is to show that 
the evolution of the growing modes of the
axionic seeds will be significantly  influenced by the breaking of the
isotropy with quite unexpected consequences.

Indeed in the case of
four dimensional (isotropic) string cosmological metrics we have that
the frequency dependence of the (logarithmic) energy spectra
associated with the axion is typically ``red'', i.e. the energy
density of axions (in critical units) increases at large
distances. This occurrence simply indicates that if we wait long enough
time these fluctuations may become dominant (possibly over-closing the
Universe) in the far future. If we go to  the case of ten
dimensional  exact solutions of the low energy beta
functions (of the type discussed in \cite{1,2}) with four expanding 
and six
contracting dimensions  flat spectra of axionic fluctuations are
allowed \cite{4b,4c} (provided the axionic fluctuations along the
internal dimensions are frozen).
 
The question we want to address is then the following: {\em if we consider
the completely anisotropic case are there any indications
 (already in four dimension) for different spectral behaviors?} 
In this sense our investigation complements and extends previous studies 
on the subject.

In order to study this problem it is useful to review very briefly
the standard results. Consider a four dimensional (isotropic)
metric of pre-big-bang type \cite{1}. 
Then, the Kasner-like nature of the solutions \cite{1}
implies that the curvature scale and the dilaton coupling are both
growing at the same rate:
\begin{equation}
a(\eta) = \biggl[-\frac{\eta}{\eta_1}\biggr]^{-\frac{1}{\sqrt{3} + 1}},
~~~~e^{\frac{\phi}{2}} = 
\biggl[-\frac{\eta}{\eta_1}\biggr]^{-\frac{\sqrt{3}}{2}}
\label{1.1}
\end{equation}
where $\phi$ is the dilaton field, and $\eta$ is the conformal time coordinate.
In this background the (canonically normalized) axionic fluctuation
evolve according to 
\begin{equation}
{\cal C}_{k}'' + \biggl[ k^2 - \frac{{\cal G}''}{{\cal G}}\biggr] {\cal C}_{k}
=0,~~~{\cal C}_{k} = {\cal G} \psi_{k},~~~{\cal G} = a ~e^{\frac{\phi}{2}}
\label{1.2}
\end{equation}
( $\psi$ is the axion field and the prime denotes derivation with
respect to conformal time coordinate; $k$ denotes the comoving momentum).
 Substituting into Eq. (\ref{1.2}) the
background given in Eq. (\ref{1.1}) we find that the solutions of 
Eq. (\ref{1.2}) can be expressed as linear combination of Bessel
functions \cite{abr} with index $\mu = \sqrt{3}$. It can be checked
that in the limit when a given mode is outside the horizon
(i.e. $\eta\rightarrow 0_{-}$) the small argument limit of the Bessel
functions leads to a growing and to a decreasing mode solutions
\begin{equation}
{\cal C}_{k}(\eta) \simeq c_1(k) \biggl[-\frac{\eta}{\eta_1}\biggr]^{\sqrt{3}
+\frac{1}{2}} + c_2(k) \biggl[-\frac{\eta}{\eta_1} \biggr]^{-\sqrt{3}+
\frac{1}{2}}
\label{1.2b}
\end{equation}
 ($c_1(k)$ and $c_2(k)$ are two integration constants fixed by the quantum
 mechanical normalization of the canonical mode function).

If the dilaton driven phase is followed by a radiation dominated phase
it is possible to compute the logarithmic energy spectrum according to
the standard techniques developed in the case of metric and gauge
fluctuations \cite{3,3b} with the result that the axion energy
density $\rho_{\psi}$ will be given, in critical units, as 
\begin{equation}
\Omega_{\psi}(\omega, \eta) = \frac{1}{\rho_{c} }\frac{d \rho_{\psi} }
{d\log {\omega}} = g_{1}^2 \Omega_{\gamma}(\eta)
\biggl[\frac{\omega}{\omega_{1}}\biggr]^{ 3 - 2 \mu}
\label{1.3}
\end{equation}
($\omega= k/a $  is the physical momentum; $\omega_1 = k_1/a \simeq
(\eta_1 a)^{-1}$ is the maximal amplified frequency;
 $g_1 = 0.1-0.01$ is the value of the coupling constant at 
the moment of the
transition from the dilaton-driven phase to the radiation dominated
phase, and $\Omega_{\gamma}(\eta)$
is the critical fraction of energy density stored in radiation at a
given time $\eta$).
From Eq. (\ref{1.3}) we see that if $\mu = \sqrt{3}$, as implied by
the background (\ref{1.1}), the spectral slope
in Eq. (\ref{1.3}) is $3 - 2 \mu = - 0.46$, namely the logarithmic
energy spectrum is decreasing in frequency. Usually these type of
spectra are called red simply because the largest power is stored at
large distance scales. If one wants to use axionic seeds for the
explanation of large scale (temperature) fluctuations in the Cosmic
Microwave Background Radiation (CMBR) red spectra do not seem ideal
\cite{4b} since one would need either flat or ``blue'' spectra (i.e. $3
- 2 \mu \gaq 0$). As noted in \cite{4b,4c} by adding six (internal)
contracting dimensions and by assuming that axion fluctuations do not
propagate along the internal dimensions, it is possible to achieve
$\mu = 3/2$. In this case the scale factors of the external, 
$3$-dimensional
(expanding) and internal, $6$-dimensional
 (contracting) manifolds will evolve, 
respectively as $a_{e}(\eta) \sim
9-\eta)^{-1/4} $ and $a_{i}(\eta) \sim (-\eta)^{1/4}$ while the dilaton is
constant \cite{1,2}. This kind of background could also be
dynamically justified not only in terms of vacuum solutions but also
in terms of a string driven dynamics where the (vacuum) background
equations are supplemented by (effective) string sources with 
negative pressure \cite{10}.

Are the conclusions of the present analysis valid in the case of a
{\em completely anisotropic four-dimensional metric}? This is the main
point we would like to investigate. 

Thus, 
we should study which is the evolution of the (super-horizon) 
axion fluctuations when we drop the assumption of isotropy 
for the four dimensional metric. We will also have to compute
precisely the spectral energy density when the axionic
inhomogeneities, amplified during the anisotropic (dilaton-driven)
phase, re-enter during an isotropic radiation dominated phase.

The plan of our paper in then the following. In Section II we will
review the basic equations describing the dynamics of anisotropic
dilaton-driven phases. We will also set up the main evolution
equations of the axionic mode functions. In Section III we will study
the evolution of axionic perturbations in the anisotropic models and
we will show, in some analytic examples, how blue energy spectra
appear. In Section IV we will concentrate our attention on
numerical solutions. Finally in Section V we will study the influence
of string tension corrections on our picture paying special attention
to the class of models leading to blue axion (logarithmic) energy
spectra. Section VI contains some final remarks and speculations.

\renewcommand{\theequation}{2.\arabic{equation}}
\setcounter{equation}{0}
\section{Basic Equations}
The low energy string theory effective action
in a four-dimensional background is
\begin{equation}
S= - \frac{1}{2\lambda^2_{s}}
\int d^4 x ~\sqrt{-g} ~e^{-\phi} \biggl[ R + g^{\alpha\beta}
\partial_{\alpha}\phi\partial_{\beta} \phi 
- \frac{1}{12} H_{\mu\nu\alpha}H^{\mu\nu\alpha}\biggr],
\label{action}
\end{equation}
where $H^{\mu\nu\alpha}$ is the antisymmetric tensor field and
$g_{\mu\nu}$ is a four-dimensional (spatially flat) anisotropic metric
\begin{equation}
g_{\mu\nu} ={\rm diag}\bigl[1, -a^2(t), -b^2(t), -c^2(t)\bigr].
\label{metric}
\end{equation}
We assume the  internal compactification radii
 to be frozen since we want to analyse,
specifically, the breaking of isotropy in the external manifold.
In order to simplify our discussion we will also restrict our
 attention to the case where $b(t) = c(t)$. This restriction is not
  essential and all the results discussed in the present paper
 can be extended to the case of $b(t)\neq c(t)$. At the same time we
 find this simplification useful for the physical intuition. Notice,
 moreover, that if $b(t)=c(t)$ the tensor modes of the geometry
 propagating along the $x$ axis can be decoupled from the scalar modes
 (this is not the case if $b(t)\neq c(t)$) \cite{6,7}. 

The components of the
 antisymmetric tensor field $H^{\mu\nu\alpha}$ can be re-written in
 terms o a pseudo-scalar axion as 
\begin{equation}
H^{\mu\nu\alpha} =
e^{\phi}~\frac{\epsilon^{\mu\nu\alpha\rho}}{\sqrt{-g}}
~\partial_{\rho}\psi.
\end{equation} 
With this choice the action (\ref{action}) becomes
\begin{equation}
S= - \frac{1}{\lambda^2_{s}}
\int d^4 x ~\sqrt{-g} ~e^{-\phi} \biggl[ R + g^{\alpha\beta}
\partial_{\alpha}\phi\partial_{\beta} \phi 
- \frac{1}{2} e^{2 \phi} g^{\alpha\beta}
\partial_{\alpha}\psi\partial_{\beta}\psi
 \biggr].
\label{action2}
\end{equation}
We want to study the situation where the axion field is not a source
of the background but its fluctuations get excited by the coupled
evolution of the metric and of the dilaton.
By varying the action with respect to the dilaton field and with
respect to the metric we obtain, after linear combinations,
 the tree-level evolution equations
\cite{1} for an anisotropic string cosmological model 
\begin{eqnarray}
&&\dot{\overline{\phi}}^2 - 2 \ddot{\overline{\phi}} + H^2 + 2 F^2 =0
\nonumber\\
&&\dot{\overline{\phi}}^2 - H^2 - 2 F^2=0
\nonumber\\
&&\dot{H} = H \dot{\overline{\phi}}, ~~~\dot{F} = F
\dot{\overline{\phi}}
\label{syst1}
\end{eqnarray}
where $\dot{\overline{\phi}}=\dot{\phi} - H - 2 F$ and the over-dot
denotes derivation with respect to the cosmic time coordinate $t$.
The dilaton driven (vacuum) solutions of this system can be written as
\begin{equation}
a(t) = \biggl[-\frac{t}{t_1}\biggr]^{\alpha},~~~ b(t) = 
\biggl[-\frac{t}{t_1}\biggr]^{\beta},~~~\phi(t) = ( \alpha + 2 \beta
-1)\log{\biggl[-\frac{t}{t_{1}}\biggr]},
\label{solt}
\end{equation}
with
\begin{equation}
\alpha^2 + 2 \beta ^2 =1.
\label{sum}
\end{equation}
This last equation implies a Kasner-like relation for the
exponents (notice that in the true Kasner case we also would have the 
condition $\alpha + 2 \beta =1$ which is not mandatory, in our case,
in order to have a solution of the system given in Eqs. (\ref{syst1})).
Notice that the solutions given in Eq. (\ref{syst1}) have in general
two branches connected by scale factor duality \cite{10b}. We will be
mainly concerned, in our analysis, with the dilaton-driven branch
where both the scale factors expand and we will comment, when
required, about other (possibly relevant) phenomenological situations.
 
For future convenience we can  express the actual
solutions in conformal time coordinate (related to the cosmic time
coordinate $t$ by the usual relation $a(\eta) d \eta = d t$) :
\begin{equation}
a(\eta) = \biggl[-\frac{\eta}{\eta_1}\biggr]^{\frac{\alpha}{1 -
\alpha}}, ~~~  b(\eta) =
\biggl[-\frac{\eta}{\eta_1}\biggr]^{\frac{\beta}{1 - \alpha}},
~~~\phi(\eta) = \frac{( \alpha + 2 \beta -1)}{1 - \alpha}
\log{\biggl[-\frac{\eta}{\eta_{1}}\biggr]}.
\label{solc}
\end{equation}
By varying the action (\ref{action2}) with respect to $\psi$ we can
obtain the  evolution equation of the axionic inhomogeneities
\begin{equation}
\partial_{\alpha}\biggl[e^{\phi}~\sqrt{- g}~ g^{\alpha\beta}
\partial_{\beta} ~\psi~\biggr] =0
\end{equation}
which can be also translated in conformal time 
\begin{equation}
\psi'' + ( \phi' + 2 {\cal F} ) \psi' - \nabla^2_{x} \psi -
\frac{a^2}{b^2} \biggl[ \nabla^2_{y} + \nabla^2_{z} \biggr] \psi =0
\label{psieq}
\end{equation}
where ${\cal F} = (\log{b})'$. Defining the canonical field as 
${\cal C} = e^{\phi/2} b~ \psi$ and going to Fourier space we can
further modify Eq. (\ref{psieq}) as 
\begin{equation}
{\cal C}_{k}'' +\biggl[ k^2_{L} + k^2_{T}\frac{a^2}{b^2} -
\frac{{\cal G}''}{{\cal G}}\biggr] {\cal C}_{k} = 0,~~~{\cal G} = 
b e^{\frac{\phi}{2}}
\label{can1}
\end{equation}
( $k_{L}$ is the modulus of the 
longitudinal momentum and $k_{T} =
\sqrt{k_{y}^2 + k_{z}^2} $ is modulus of the transverse momentum). It
is clear that this equation reduces to Eq. (\ref{1.2}) if the
background is completely isotropic. In fact taking $a\rightarrow b$
and recalling that $k= \sqrt{k^2_{L} + k^2_{T}}$, we have
that Eq. (\ref{1.2}) is recovered. At the same time we can clearly see
that if $a\neq b$ and the background is not isotropic Eq. (\ref{1.2})
differs qualitatively from Eq. (\ref{can1}) which 
can also be written in a slightly different form in
terms of $k$ and of the ratio of the transverse and longitudinal momentum $r=
k_{T}/k_{L}$:
\begin{equation}
{\cal C}_{k}''
 +\biggl[ \frac{k^2}{1 + r^2} \biggl( 1 + r^2 \frac{a^2}{b^2} \biggr)-
\frac{{\cal G}''}{{\cal G}}\biggr] {\cal C}_{k} = 0
\label{can2}
\end{equation}
In terms of the rapidity $y= (1/2)\log{[(k+k_{L})/(k-k_{L})]}$, 
$r= (\sinh{y})^{-1}$.
This form of the evolution equation will be useful  in Section II for
the study of the axionic (canonical) mode function
${\cal C}_{k}$ outside the horizon.
Inserting the anisotropic dilaton-driven solution of Eqs. (\ref{sum}) and
(\ref{solc}) into Eq. (\ref{can1}) we obtain 
\begin{equation}
{\cal C}_{k}'' +\biggl\{ k^2_{L} + 
k^2_{T}\biggl[- \frac{\eta}{\eta_1}\biggr]^{\gamma} -
\frac{\mu^2 -\frac{1}{4}}{\eta^2}\biggr\} {\cal C}_{k} = 0, 
\label{can3}
\end{equation}
where 
\begin{equation}
\gamma= \frac{2(\alpha - \beta)}{1-\alpha},~~~
2\mu = |2 \lambda - 1|,~~~\lambda= \frac{ \alpha + 4 \beta -1}{ 2(1 -
\alpha)}
\label{ab}
\end{equation}
Before dealing with the analytical and numerical solutions of Eq. 
(\ref{ab}) we want to present some more qualitative considerations.
Comparing  Eq. (\ref{1.2}) to Eq. (\ref{can3}) we can notice some
analogies but also crucial differences. Both equations can be seen
as ``Schroedinger-like'' equations but with different ``potentials''.
More specifically, one can find a similarity  between Eq. (\ref{can3}) and 
the radial part of the
Schroedinger equation in a spherically symmetric 
 potential  where the term going as
 $\eta^{-2}$ would correspond to the (orbital) angular momentum.
The term containing $k_{T}$ would correspond instead to the (central)
potential \cite{abr,lan}.
In our problem there are two relevant cases: either $\gamma <-2 $ or
$\gamma >- 2$. If $\gamma<-2$, then the ``potential'' term might
become larger, in the limit $\eta\rightarrow 0_{-}$ (i.e. axion
fluctuations outside the horizon), than the ``angular momentum'' term.
We can immediately see, amusingly enough, that this situation is never
realized for the Kasner-like solutions defined by Eq. (\ref{sum}). In
fact the case $\gamma <-2$ corresponds, in the $(\alpha,\beta)$ plane
to $\beta > 1$ which is not realized since in the (anisotropic) vacuum
solutions $\alpha$ and $\beta$ have to lie on the ellipse $\alpha^2 +
2 \beta^2=1$, and, therefore we will always have that $|\beta |\leq
1/\sqrt{2}\sim 0.707...$. Thus, we could always expect the $\eta^{-2}$
term to be, at some point, dominant for $\eta \rightarrow 0_{-}$. 
\begin{figure}
\centerline{\epsfxsize = 7 cm  \epsffile{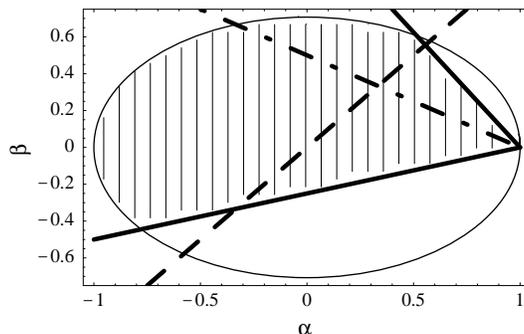}} 
\caption[a]{We illustrate the interplay between the solutions given in
Eqs.  (\ref{solt})-(\ref{solc}) and the behavior of the axionic mode
function outside the horizon. The ellipse is the geometrical version of
Eq. (\ref{sum}). All the $\alpha$ and $\beta$ on the ellipse give rise
to solutions of the low energy beta functions in the anisotropic
metric of Eq. (\ref{metric}) with $c = b$. 
The arc bounded by  the two full
(thick) lines defines the class of anisotropic models leading to
$\mu \leq 3/2$. The shading is only meant to guide the eye. The
intersections of the full lines with the ellipse define the models for
which $\mu = 3/2$. The dashed curve correspond to $\beta =
\alpha$. For $\beta >\alpha$ (above the dashed line) 
we have, in Eq. (\ref{can3}) that $\gamma<0$. The dot-dashed line
 corresponds to the case where $\lambda
= 1/2$ (see Eq. (\ref{ab})), implying $\mu =0$. }
\label{f1}
\end{figure}
In order to visualize
our qualitative arguments, let us look at Fig. \ref{f1} where
the ellipse is just the plot (in the ($\alpha,\beta$) plane) of
Eq. (\ref{sum}) defining the Kasner-like solution of Eqs. (\ref{syst1})
and (\ref{solc}). The dashed line corresponds to $\beta = \alpha$. If
$\beta > \alpha$ (above the dashed line) $\gamma<0$ and the
``potential'' term goes to zero for $\eta \rightarrow -\infty$. 
Since $\eta^{-2}$ becomes dominant for $\eta \rightarrow 0_{-}$, the
solutions of Eq. (\ref{can3}) should always go to Bessel functions
with index $\mu$.  

In the {\em anisotropic} case the index
$\mu$ can be different from the Bessel index appearing in
Eq. (\ref{1.2}) for the {\em isotropic} case. Given that the (logarithmic)
 energy
spectrum is determined by  the dominant solution of
Eq. (\ref{can3}) in the limit $\eta\rightarrow 0_{-}$, we can expect
for $\mu \laq 3/2$ a flat or blue logarithmic energy spectrum. In the
$(\alpha,\beta)$ plane the region corresponding to $\mu \laq 3/2$ can
be obtained from Eq. (\ref{ab}) with some elementary algebra. The
result is reported in Fig. \ref{f1}. The arc of the ellipse bounded by
the two (full) thick lines correspond to (anisotropic) vacuum solutions
leading to growing modes compatible with ``blue'' spectra. The
intersections between the two full lines and the ellipse correspond to
the case $\mu=3/2$ (flat spectrum). Of course not all the models
within the full lines are realistic. 

For examples we could have, in
principle, growing spectra with $\alpha>0$ and $\beta>0$ (first
quadrant). These models are clearly {\em not} realistic since they
would describe a dilaton-driven contraction more than a dilaton-driven
expansion. More realistic are the models in the second quadrant where
one of the two scale factors expands and the other contracts. 

Finally
in the third quadrant we have the most physical situation where the
two scale factors {\em both expand}, though at different rates. It is amusing
to notice that the intersection between the full line and the ellipse
picks up a dilaton-driven solution with $\mu= 3/2$ and expanding scale
factors. The lower full line has equation $4 \beta = (\alpha-1)$ and
the intersection we are interested occurs for $\alpha = - 7/9$ and
$\beta = - 4/9$. Going to Eq. (\ref{solc}) we see that, in this case
$\phi(\eta) \sim -3/2 \log{[- \eta/\eta_1]}$. 
In this case, on the
basis of the asymptotic behavior of Eq. (\ref{can3}) in the small
$|\eta|$ limit, the
solution outside of the horizon will be given by
\begin{equation}
{\cal C}_{k}(\eta) \simeq c_{1}(k) \biggl[-\frac{\eta}{\eta_1}\biggr]^{-1} +
c_{2}(k)  \biggl[-\frac{\eta}{\eta_1}\biggr]^{2}.
\end{equation}
which coincides with the solution one would obtain in the
ten-dimensional case where the external manifold has scale factor 
$a_{e}(\eta) \sim (-\eta)^{-1/4}$ and the internal one has scale
factor $a_{i} \sim(- \eta)^{1/4}$. 

Another interesting case is represented by  the model 
$\alpha = -1$ and $\beta =0$. In this case $\phi(\eta) \sim -
\log{[-\eta/\eta_1]}$. 
This solution might not
seem fully realistic and indeed we will use it as a toy model. At the
same time one can argue that this type of solutions might be the result, in the
pre-big-bang scenario, of spherically symmetric initial conditions which are
asymptotically trivial in the past \cite{5}. 
It is quite amusing to work out the Einstein frame
description of this solution. The Einstein frame scale factors are
related to the ones of the String frame by trivial conformal
rescalings involving the dilaton, namely 
\begin{equation}
\tilde{a} = e^{ - \frac{\phi}{2}} a, 
~~~ \tilde{b} = e^{ -\frac{\phi}{2}} b.
\end{equation}
Since the conformal time coordinate does not change passing from one
frame to the other (i.e. $d \eta \equiv d\tilde{\eta}$ we can obtain
easily that, in the Einstein frame, our solution looks like $\tilde{a}
\sim {\rm const.} $ and $\tilde{b} \sim \sqrt{-t}$ (since 
$\tilde{a}(\eta)$ is constant $t$ and $\eta$ coincide up to a constant
factor). 

It is interesting to point out that this solution was
Kasner-like in the String frame (i.e. $\alpha^2 + 2 \beta^2 =1$) and
it becomes true Kasner in the Einstein frame.
This solution is well known in the theory of the
gravitational collapse  \cite{11} and it
might be thought as the limit (in the vicinity of the singularity)
of a spherically symmetric space filled with a stiff (perfect) fluid.
It was indeed recently stressed that the Einstein frame
description dilaton-driven cosmologies in four dimensions is closely
connected with the problem of the gravitational collapse of a stiff
fluid and that, consequently, string cosmological models of
pre-big-bang type are as fine-tuned as the gravitational collapse
\cite{5}. 

If $\alpha = -1 $ and $\beta =0$ we have that in Eq. (\ref{can3})
$\gamma= -1$. It is amusing to notice that in this case the
Eq. (\ref{can3}) is analytically solvable and it is (formally)
equivalent to the Schroedinger equation in a Coulomb potential 
with complex
electromagnetic coupling. We will use this analogy in the next
section. For the moment we can
guess that outside the horizon the solution to Eq. (\ref{can3}) will
be well approximated by
\begin{equation}
{\cal C}_{k} = c_1(k) \biggl[ - \frac{\eta}{\eta_1}\biggr]^{\frac{3}{2}} + 
 c_2(k)  \biggl[ - \frac{\eta}{\eta_1}\biggr]^{- \frac{1}{2}}.
\label{bl0}
\end{equation}
Again this solution was simply obtained by solving Eq. (\ref{can3}) in
the small $\eta$ limit where the ``Coulomb'' contribution can be
neglected. Since this is nothing but the small argument limit of
Hankel functions with $\mu =1$, we again argue that the exact solution
should give us a blue (logarithmic) energy spectrum of the type of the
one given in Eq. (\ref{1.2}) with spectral index $ 3 - 2 \mu = 1$.

\renewcommand{\theequation}{3.\arabic{equation}}
\setcounter{equation}{0}
\section{Blue Spectra}

In some cases, analytical
solutions of Eq. (\ref{can3}) can be found. 
We start to discuss blue spectra and, more precisely we analyze the
model mentioned in the previous Section with $\alpha =-1$. 
In this case Eq. (\ref{can3}) becomes, using $k$ and $r$
\begin{equation}
{\cal C}_{k}'' + \biggl\{ \frac{k^2}{1 + r^2}\biggl[1  + r^2  
\biggl(-\frac{\eta}{\eta_{1}}\biggr)^{-1} \biggr] - \frac{3}{4\eta^2}
\biggr\}
{\cal C}_{k} =0.
\label{bl1}
\end{equation}
This equation looks indeed very similar to the (radial) Schroedinger
equation for the Coulomb problem. In order to solve Eq. (\ref{bl1}) we
have to match the standard notations appearing in the theory of
special functions \cite{abr}.
Defining  a new (complex) time as 
\begin{equation}
\tau = \frac{2 i k \eta}{ \sqrt{1 + r^2}}
\end{equation}
we can  re-write Eq. (\ref{bl1}) as
\begin{equation}
\frac{d^2 {\cal C}_{k}}{d \tau^2} + \biggl[ -\frac{1}{4} +\frac{
\zeta}{\tau} - \frac{3}{4 \tau^2}\biggr] {\cal C}_{k} =0
\label{bl2}
\end{equation}
with $\zeta =  (i/2) \epsilon r$. It is  helpful to
define $\epsilon = k_{T} \eta_1= k \eta_1 r/\sqrt{1 + r^2}$ purely
for algebraic reasons. Eq. (\ref{bl2}) is one of the well known forms
of the Whittaker's equation \cite{abr} whose solutions are the
corresponding Whittaker's functions, related to confluent
hypergeometric functions. We are interested in
solutions which represent incoming axionic waves for $\eta
\rightarrow - \infty$ and then our solution will have the form
\begin{equation}
W_{k}(\tau) = e^{- \frac{\tau}{2} } \tau ^{\frac{3}{2}}U[ \frac{3}{2} -
\frac{i}{2} \epsilon r, 3, \tau]
\end{equation}
where $U[m, n, \tau]$ is the (Kummer) confluent hypergeometric
function with parameters $m$ and $n$. Therefore the (properly normalized) form of the solution of
Eq. (\ref{bl1}) is 
\begin{equation}
{\cal C}_{k}(\eta) = \frac{1}{\sqrt{k}}\sqrt{\frac{2}{\pi}} 
e^{ - i \frac{ k \eta}{\sqrt{1 +
r^2}} + i\frac{\pi}{4}} \biggl[ \frac{ 2 i k \eta }{\sqrt{1 +
r^2}}\biggr]^{\frac{3}{2}} U [ \frac{3}{2} - \frac{i}{2} r \epsilon, 3,
2 i \frac{ k \eta}{\sqrt{r^2 + 1}}].
\label{sol}
\end{equation}
We can immediately notice, as expected, that the small argument limit
of the previous solution (i.e. $k\eta \ll 1$) exactly reproduces the
asymptotic evolution obtained in Eq. (\ref{bl0}). 

Having derived the solution of Eq. (\ref{bl1}) we can also compute the
energy spectrum of the axions amplified by the transition from an
anisotropic dilaton-driven solution to a (completely isotropic) radiation
dominated epoch occurring at $\eta = - \eta_1$.
For $\eta > - \eta_1$ the evolution equation of the axion fluctuations
will be given by 
\begin{equation}
{\cal C}_{k} '' + k^2 {\cal C}_{k} =0,
\label{bl3}
\end{equation}
because the dilaton coupling freezes, after $-\eta_1$, to a constant
value. A similar calculation was  performed in  \cite{4b} under
the assumption that the isotropisation of the geometry occurred prior
to the onset of the dilaton-driven phase (indeed described,
in \cite{4b} only by one scale factor). 

The frequency mixing coefficient $c_{-}(k)$, determining the spectral
number of produced axions, can be explicitly computed by matching, at
$\eta = - \eta_1$ the general ``outgoing'' solution of Eq. (\ref{bl3})
\begin{equation}
{\cal C}_{k}(\eta) = \frac{1}{\sqrt{k}}\biggl[ c_{+} e^{ - i k (\eta +
\eta_1) } + c_{-}  e^{  i k (\eta +\eta_1) }\biggr],~~~\eta>-\eta_1,
\label{free}
\end{equation}
to the ``incoming'' solution (\ref{sol}) of Eq. (\ref{bl1}) which
represents positive frequency modes in the $\eta\rightarrow - \infty$
limit (notice, however, that the incoming solution does not define,
asymptotically a Bunch-Davies vacuum because of the presence of the
term ``Coulomb'' term in Eq. (\ref{bl1}) which 
is dominant in the far past).

Following this procedure we determined the mixing coefficients
reported in Appendix A. The important point to mention is that for 
$k\eta_1 >1 $ the procedure leading to the mixing coefficients is no
longer valid but  the mixing coefficient can be anyway estimated by
using in Eq. (\ref{bl1}) a smooth ``potential'' interpolating between 
$\eta <-\eta_1$ and $\eta>-\eta_1$. For $k \eta_{1}>1$ the mixing
coefficients are indeed exponentially suppressed \cite{12} and can be
ignored for the purpose of this paper. Therefore, taking the limit
(for $k\eta_1 <1$) of the results reported in Appendix A  we have
\begin{equation}
|c_{-}(x_1)| \simeq  \frac{[S(r)]^{-1/2}}{4 \pi} x_{1}^{- 3/2}
\label{bog1}
\end{equation}
where $S(r) = 1/\sqrt{r^2 + 1} $ and $x_1 = k\eta_1$.

The spectral energy density
\begin{equation}
\rho_{\omega} =  \frac{ d \rho_{\psi}}{d \log{\omega}} \simeq
\frac{\omega^4}{\pi^2}  |c_{-}(\omega)|^2
\end{equation}
is the variable frequently adopted to characterize the distribution of
the produced particles \cite{2,3,4b}, for our case reads
\begin{equation}
\Omega_{\psi} (\omega,\eta) = \frac{\rho_{\omega}}{\rho_{c}}=
\frac{g_1^2 \Omega_{\gamma}(\eta) }{16 \pi^4 S(r)} 
\biggl(\frac{\omega}{\omega_1}\biggr)
\label{om}
\end{equation}
where $\rho_{c}$ is the critical energy density and
$\Omega_{\psi}(\omega, \eta)$ is the (critical) fraction of produced
(mass-less) Kalb-Ramond axions. Notice that $n(\omega)=
|c_{-}(\omega)|^2$ can be interpreted, in a fully quantum mechanical
treatment of the processes of cosmological particle production
\cite{2,13}, as the mean number of axions with physical momentum 
$\omega$. In Eq. (\ref{om}) $g_1 = e^{\phi(\eta_1)/2} \sim 0.1-0.01 $
is the final value of the string coupling parameter 
at the time of the transition to the radiation dominated epoch. In
terms of $g_1$ the string and Planck mass are related as $M_{s} = g_1
M_{P}$. 

The spectral slope appearing in Eq. (\ref{om}) is exactly the
one we guessed in Section II using the approximate solution of
Eq. (\ref{can3}) in the small $\eta$ limit for the axion mode
function. The result is that
 the spectral energy density is only slightly tilted towards
large frequencies. We remind, incidentally,
that in string cosmological models blue spectra are also
possible for gauge fields \cite{3b}, but not for metric fluctuations 
\cite{2}. In the case of gravitons, for example, the typical slope for
the spectral energy density is close to 3 (``violet'' energy spectra).

Thus  also in the fully anisotropic case
the spectral distribution of the amplified axions is determined by the
solution of the mode equation in the small $\eta$ limit. Furthermore,
on the basis of the analytical treatment performed in the present
Section we can argue that as  long as $\gamma > - 2 $ in
Eq. (\ref{can3}), the spectral slope of the spectral energy density
$\rho_{\omega}$ is completely determined by the ``angular momentum''
term appearing in Eq. (\ref{can3}). On the other hand the spectral
amplitude does depend upon the details of the particular anisotropic
model and upon the relative weight of the longitudinal and transverse
momentum (i.e. $r = k_{T} /k_{L}$) 
during the anisotropic phase. In particular  if
$r\rightarrow 0$ (i.e. $k_{T} \ll k_{L}$) $S(r)
\rightarrow 1$. In the
large $r$ limit (i.e. $k_{T} \gg k_{L}$) $S(r) \sim r$.
 From a physical
point of view, $r\sim 1$ implies a fairly isotropic (initial) momentum
distribution and it makes sense only when $a(\eta)\rightarrow
b(\eta)$. Conversely, $r$ very different from 1 implies either the
dominance of transverse momenta or the dominance of 
the longitudinal momenta. We see that $r$ cannot be arbitrarily large
since we have always to demand 
$\Omega(\omega,\eta) <1$. Imposing the critical energy bound on
the axion fluctuations we get, indeed, that $r< 16 \pi^4/g_1^2 $
(i.e. $r< 10^{5}$ if $g_1 = 0.1$). 

\begin{figure}
\centerline{\epsfxsize = 7 cm  \epsffile{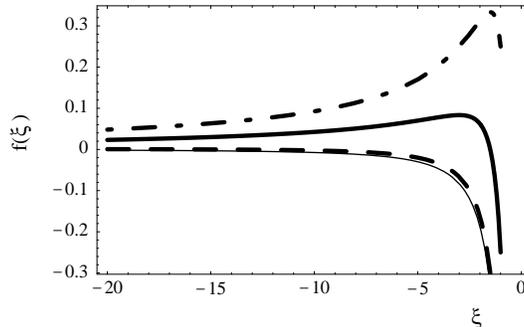}} 
\caption[a]{We plot different regimes of Eq. (\ref{pot}). With the
thick (full) line we have the case $x_1=1$ and $r=1$. With the dashed
line we illustrate the case $x_1 = 0.1$, $r=1$, whereas with the
dot-dashed line we illustrate the case $x_1 = 1$ and $r =
10^{3}$. Notice that the thin (full) line is for the case $x_1=
0.001$, $r=1$.}
\label{f2}
\end{figure}

One can understand, more physically, the $r$ and $k\eta_1$ dependence 
by looking more closely at Eq. (\ref{bl1}). Let us write the
``Coulomb'' potential and the ``angular momentum'' in a rescaled form,
namely
\begin{equation}
f(\xi, x_1, r) =\frac{x_1^2 ~r^2}{r^2 + 1} \frac{1}{\xi} - \frac{3}{4 \xi^2}
\label{pot}
\end{equation}
where $\xi = [-\eta/\eta_1]$ and, as usual, $x_1 = k \eta_1$.
The plot of the function (\ref{pot})  
for (different values of the parameters) is reported in Fig. \ref{f2}.
Since we want to consider modes which are amplified (and not
exponentially suppressed) we have to focus on the case $x_1 \laq 1$. From
Fig. \ref{f2} we learn that as soon as $x_1$ gets smaller and smaller
(i.e. frequencies smaller than the maximal) the ``Coulomb'' term in 
Eq. (\ref{pot})
gets even more negligible if compared with the term going as
$\xi^{-2}$. Moreover, for large $r$ and high frequencies (i.e. $x_1\sim
1$) the potential barrier gets higher. Thus, if we are going to
explore a frequency range $k\ll\eta_{1}^{-1}$ (i.e. $\omega\ll\omega_1$)
our approximation improves.

\renewcommand{\theequation}{4.\arabic{equation}}
\setcounter{equation}{0}
\section{Flat Spectra}

\begin{figure}
\centerline{\epsfxsize = 7 cm  \epsffile{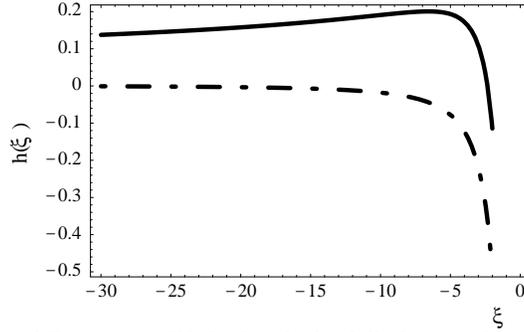}} 
\caption[a]{We illustrate different cases of Eq. (\ref{num}). With the
thick (full) line we have the case $x_1=1$ and $r=1$. With the dashed
line we plot the case $x_1 = 0.1$, $r=10$.}
\label{f3}
\end{figure}

\begin{figure}
\begin{center}
\begin{tabular}{|c|c|}
      \hline
      \hbox{\epsfxsize = 6.5 cm  \epsffile{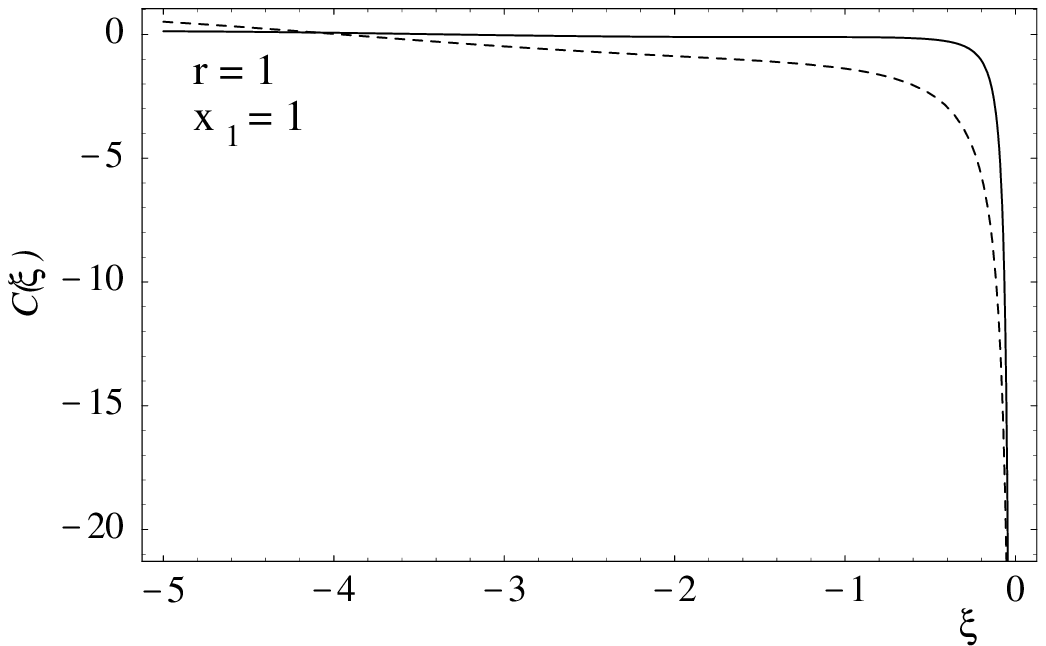}} &
      \hbox{\epsfxsize = 6.5 cm  \epsffile{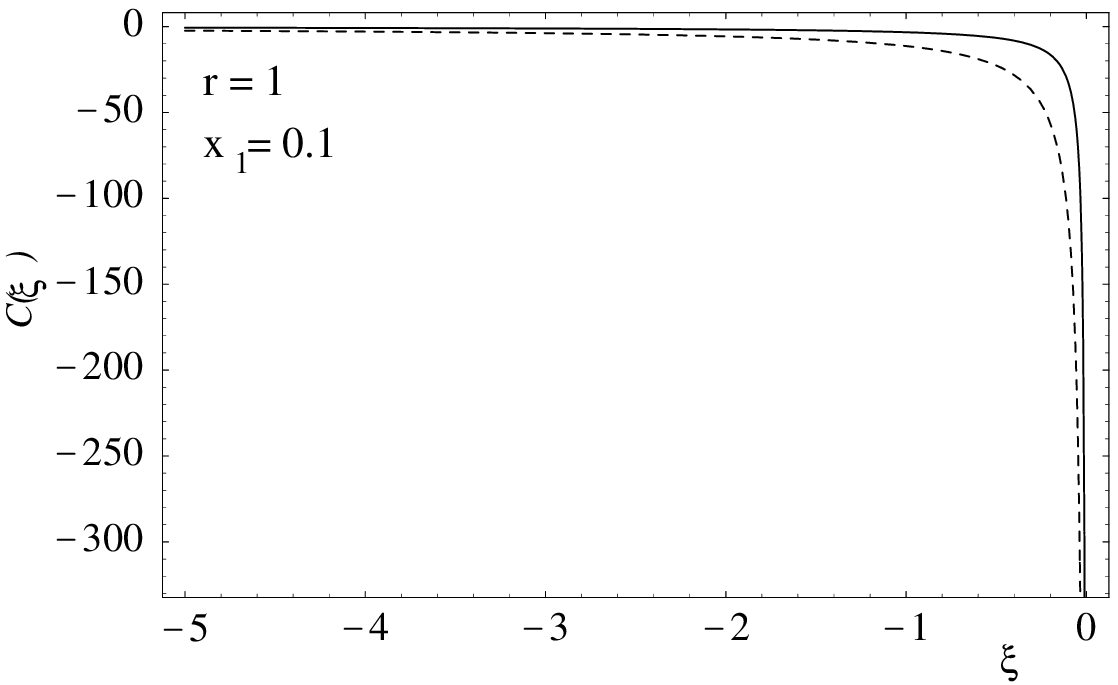}} \\
      \hline
      \hbox{\epsfxsize = 6.5 cm  \epsffile{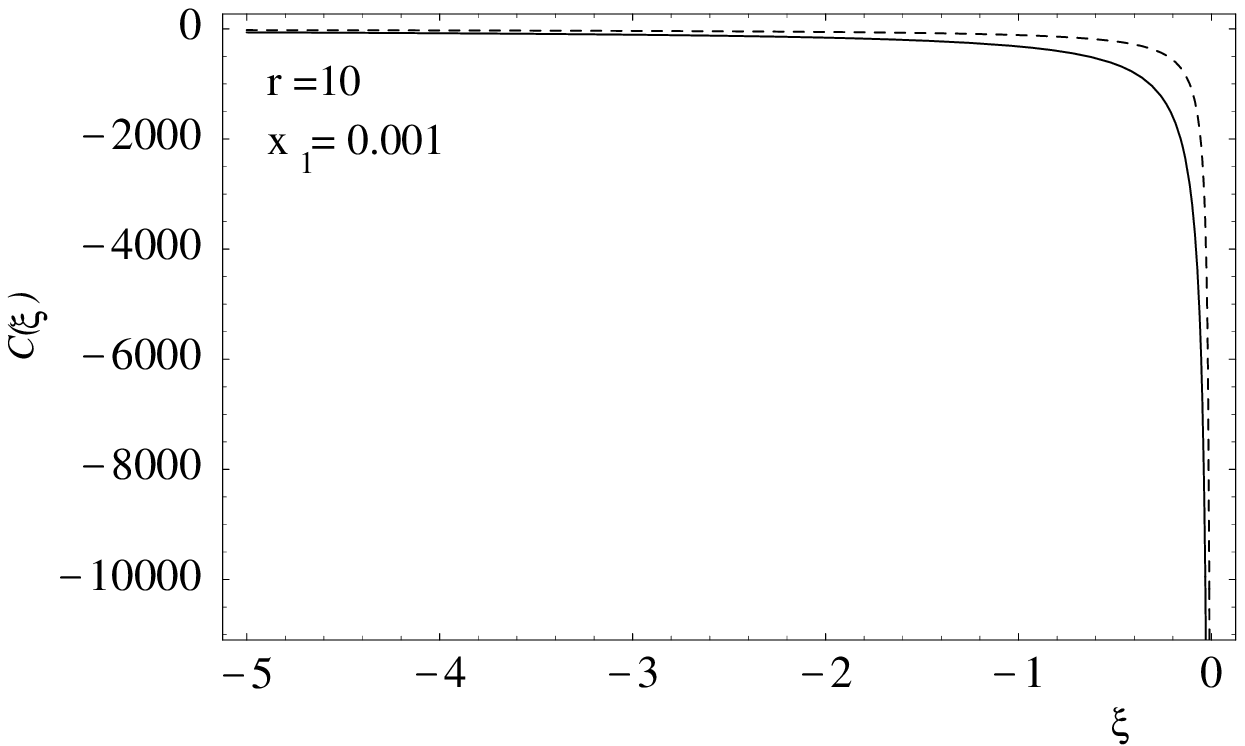}}  &
      \hbox{\epsfxsize = 6.5 cm  \epsffile{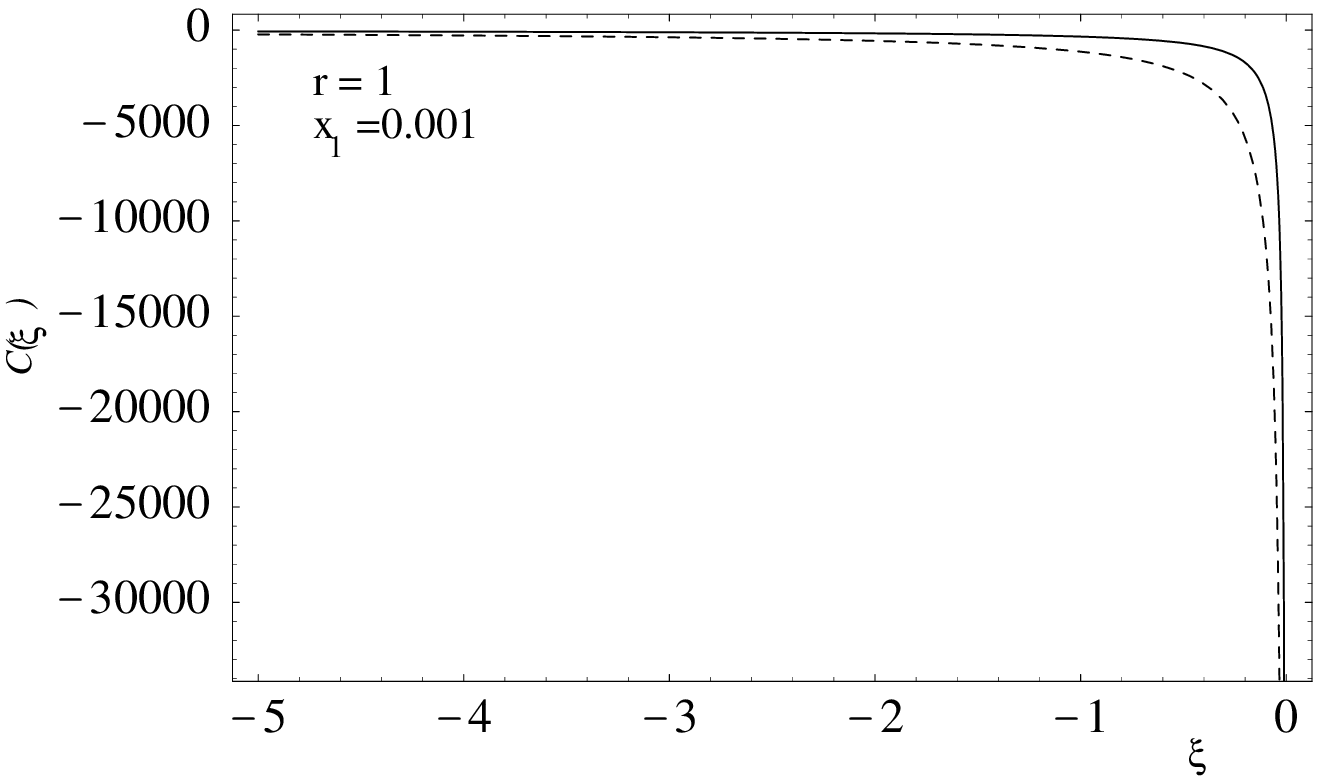}}\\
      \hline
\end{tabular}
\end{center}
\caption[a]{We report the numerical solutions of Eq. (\ref{exact}) for
decreasing values of $x_1$. We see also report the approximate
solutions (dashed curves) given in Eq. (\ref{approx}).}
\label{f4}
\end{figure}

Having discussed the case of a specific blue spectrum we want now to
treat the case of a flat spectrum in anisotropic dilaton-driven
models.  We argued in Section II that in order to obtain a flat
spectrum we have to require $\mu = 3/2$, which does correspond either
to $\alpha =1$ and $\beta = 0$ or to $\alpha = - 7/9$ and 
$\beta = - 4/9$. The case $\alpha =1$ $\beta =0$ is clearly {\em not 
realistic} : it does correspond to a contracting Universe (in the
string frame). The interesting case is indeed the second where the
dilaton growth triggers a superinflationary phase where the different
dimensions expand with different rates: 
\begin{equation}
a(t) \sim \biggl[-\frac{t}{t_1}\biggr]^{- \frac{7}{9}},~~~b(t) 
\sim \biggl[-\frac{t}{t_1}\biggr]^{- \frac{4}{9}},~~~\phi(t) = -
\frac{8}{3} \log{\biggl[- \frac{t}{t_1}\biggr]}.
\end{equation}
For this background the explicit for of Eq. (\ref{can3}) is
\begin{equation}
{\cal C}'' + \biggl\{ \frac{k^2}{1 + r^2} \biggl[ 1 + r^2 
\biggl(-\frac{\eta}{\eta_{1}}\biggr)^{-3/8}\biggr] 
- \frac{2}{\eta^2}\biggr\}{\cal C} =0,
\end{equation}
which becomes, going to dimensionless (conformal) time,
\begin{equation}
\frac{d^2 {\cal C}}{ d \xi^2} + \{ \frac{x_1^2}{ r^2 + 1} - h(\xi,
x_1, r)\} {\cal C} =0, ~~~ \xi = \frac{\eta}{\eta_1},
\label{exact}
\end{equation}
where
\begin{equation}
h(\xi, x_1, r) 
=\frac{r^2}{r^2 + 1}\frac{x_1^2}{\xi^{3/8}}  - \frac{2}{\xi^2} .
\label{num}
\end{equation} 
In Eq. (\ref{num}) the analogous of the Coulomb term goes as
$\xi^{-3/8}$. Therefore, we expect the (numerical) solutions of
 Eq. (\ref{exact})
to converge more slowly than in the case of Eq. (\ref{bl0}) 
to the solution one would
guess taking into account only the leading contribution for small $\xi$.
 This is
illustrated in Fig. \ref{f3} where the function $h(\xi)$ is reported
for different values of the parameters.
Eq. (\ref{exact}) can be solved numerically. In Fig. (\ref{f4}) the
numerical solutions of Eq. (\ref{exact})
are reported for different values of the primordial rapidity. 
For comparison (dashed lines) we also report, 
for the same values of the parameters, the behavior of the
approximate solutions of Eq. (\ref{exact}) obtained by neglecting the
term $\xi^{-3/8}$. For $x_1\laq 0.001$ the approximate solutions can
be definitely trusted in the small $\xi$ limit.
The approximate solutions plotted with the dashed line in
Fig. (\ref{f4}) are 
\begin{equation}
{\cal C}(\xi) = \frac{1}{\sqrt{k}}\sqrt{\xi x_1 S(r)}\{ J_{3/2}[
 \xi x_1 S(r)] -
i Y_{3/2} [ \xi x_1 S(r)]\}
\label{approx}
\end{equation}
By directly matching Eq. (\ref{approx}) with Eq. (\ref{free}) we
obtain that the relevant mixing coefficient is
\begin{equation}
|c_{-}|^2 \sim {\cal K}(r) \frac{1}{8 \pi^2~S(r)~x_1^4} 
\end{equation}
corresponding to 
\begin{equation}
\Omega_{\psi} (\omega,\eta) = \frac{\rho_{\omega}}{\rho_{c}}=
\frac{{\cal K}(r)}{16 \pi^4 S(r)} 
g_1^2 \Omega_{\gamma}(\eta) 
\label{om2}
\end{equation}
Notice that ${\cal K}(r)$ is a (frequency independent) 
factor which can be precisely determined numerically and which
becomes important in the limit $x_1\sim 1$. If $x_1 \laq 0.001$,
${\cal K}(r)\sim 1$.

It would be desirable to find approximate (but analytical) expressions 
describing the evolution of the mode function outside of the horizon in the 
general case of anisotropic backgrounds of the type of the one introduced in 
Eq. (\ref{solc}). 
The evolution equation (\ref{can1}) for the axionic mode function can be 
written as
\begin{equation}
{\cal C}_{k}'' + \biggl[ k^2 - V_{k}(\eta) \biggr] {\cal C}_{k} =0, 
~~~ V_{k}(\eta) =
 \frac{k^2 r^2}{ 1 + r^2}\biggl( \frac{b^2 - a^2 }{b^2}\biggr) +
 \frac{{\cal G}''}{{\cal G}}
\label{cor}
\end{equation}
The effective potential appearing in Eq. (\ref{cor}) generalizes the 
previous results of Eqs. (\ref{pot}) and (\ref{num}).
Notice that the situation of a $k$-dependent effective potential is not so 
uncommon since it arises in the case of perturbations of kaluza-Klein 
backgrounds with fluctuating internal dimensions \cite{2} and in the case 
gravitational waves  propagating in a background whose action contains higher 
derivatives \cite{k1}.
Notice also that in the isotropic limit (i.e. $ b\rightarrow a $) Eq. 
(\ref{cor}) goes to Eq. (\ref{1.2}) since ${\cal G} \rightarrow a e^{\phi/2}$
and the term proportional to the anisotropy (i.e. $(b^2 - a^2)/b^2$) exactly 
vanishes. We want to solve Eq. (\ref{cor}) in the limit $\eta 
\rightarrow 0_{-}$ and show that the correction to the leading order solution 
(obtained by dropping the $b^2 - a^2)/b^2$ term) is indeed small.
The equation we have to solve is essentially
\begin{equation}
C_{k}'' \simeq V_{k}(\eta) C_{k}.
\label{cor2}
\end{equation}
By writing the solution to Eq. (\ref{cor2}) as 
\begin{equation}
C_{k}(\eta) \sim \overline{C}_{k}(\eta) + \epsilon_{k}(\eta), 
\end{equation}
where $\overline{C}_{k}$ is the leading order solution
\begin{equation}
\overline{C}_{k} \sim A {\cal G}(\eta) + B {\cal G}(\eta) \int^{\eta} 
\frac{d \eta'}{{\cal G}(\eta')^2}
\end{equation}
we can find that $\epsilon_{k}(\eta)$ is simply given by 
\begin{equation}
\epsilon_{k}(\eta) \sim \frac{k^2 r^2 }{1 + r^2} {\cal G}(\eta) \int^{\eta} 
{\cal G}(\tau) \overline{C}(\tau) \biggl(\frac{b^2  - a^2}{b^2} \biggr)
 \int^{\tau}  \frac{d \eta''}{{\cal G}(\eta'')^2}
\end{equation}
($A$ and $B$ are arbitrary constants).
Notice, again, that the correction vanishes exactly for $b\rightarrow a$.
If we insert in the previous equations the background solutions of 
Eq. (\ref{solc}) we get  
\begin{eqnarray}
&&{\cal C}_{k}(\eta) \sim
 A  \biggl\{ \biggl(- \frac{\eta}{\eta_1}\biggr)^{\lambda} 
+ \frac{(k \eta_1)^2}{2 (\gamma +1) ( 1 - 2 \lambda) }\frac{r^2}{1 + r^2} 
\biggl[ (\gamma + 1) \biggl(- \frac{\eta}{\eta_1}\biggr)^{\lambda + 2} - 
2 \biggl(- \frac{\eta}{\eta_1}\biggr)^{ \gamma + 1 + 2 \lambda}\biggr]\biggr\}
\nonumber\\
&& + B \biggl\{ \frac{1}{1 - 2 \lambda}
\biggl(- \frac{\eta}{\eta_1}\biggr)^{1 - 2 \lambda} + \frac{(k \eta_1)^2}{ 
(1-2 \lambda)^2(3 - 2 \lambda) (3 -  \lambda + \gamma)} 
\frac{r^2 }{r^2 + 1} \biggl[ (3 - 2 \lambda + \gamma) \biggl(-
 \frac{\eta}{\eta_1}\biggr)^{3 -  \lambda} 
\nonumber\\
&&- 
(3 - 2 \lambda)\biggl(-\frac{\eta}{\eta_1} \biggr)^{3 - \lambda + \gamma}
\biggr]\biggr\}
\label{fo}
\end{eqnarray}
In each of the curly brackets the first term is the leading solution whereas
 the term proportional to $(k\eta_1)^2$ is the correction. Notice that the 
correction is subleading in the limit $\eta\rightarrow 0_{-}$. Take, for 
instance, the case investigated in this section, namely $\lambda = -1$
 and $\gamma= - 3/8$. We can easily find, from Eq. (\ref{fo}) 
\begin{eqnarray} 
{\cal C}_{k}(\eta) \sim && A \biggl\{ \biggl( - \frac{\eta}{\eta_1}
\biggr)^{-1} 
+ \frac{4}{39} (k\eta_1)^2 \frac{r^2}{r^2 + 1} \biggl[ \frac{13}{8} 
\biggl(-\frac{\eta}{\eta_1} \biggr) 
- 2 \biggl(-\frac{\eta}{\eta_1}\biggr)^{\frac{5}{8}}\biggr]\biggr\}
\nonumber\\
&+& B \biggl\{\frac{1}{3} \biggl(-\frac{\eta}{\eta_1}\biggr)^2 + 
\frac{(k\eta_1)^2}{370} 
\frac{r^2}{r^2 + 1} \biggl[\frac{37}{8} \biggl(-\frac{\eta}{\eta_1}\biggr)^{4} 
- 5 \biggl(-\frac{\eta}{\eta_1}\biggr)^{\frac{29}{8}}\biggr\}
\end{eqnarray}
Notice that the leading super-horizon contribution 
to the evolution of the axionic mode function is given, 
as expected, by the term going as $\eta^{-1}$ appearing in the first curly 
bracket. In both the curly bracket we have the term $(k\eta_1)^2$ 
which is always smaller than one since $\eta_{1}^{-1}$ is the maximal 
amplified frequency. Moreover, in both the curly brackets all the terms 
multiplied by $(k\eta_1)^2$ are vanishingly small outside the horizon.
We could repeat the same calculation by inserting different (expanding) 
anisotropic backgrounds into Eq. (\ref{fo}). We find that
 for the class of models defined on the arc of the 
``vacuum'' ellipse depicted in Fig. \ref{f1} all the corrections are inmdeed
 small and the leading contribution is given, as foreseen, by the growing mode.
Notice that we already knew that from the considerations of Section I and II. 
We the present approach we are also able to compute, analytically, the 
corrections to the leading solution.

\renewcommand{\theequation}{5.\arabic{equation}}
\setcounter{equation}{0}
\section{Anisotropy and string tension correction}
The main assumption behind the calculations we presented has been the
presence of an {\em anisotropic} string cosmological phase. In previous
calculations \cite{4b,4c} the axion spectra were actually computed
assuming that the (four dimensional) background describing the 
 dilaton-driven dynamics was completely
{\em isotropic}. Therefore, in \cite{4b,4c} it was also implicitly assumed
that the isotropization of the four (expanding) dimensions 
occurred prior to the onset of the
dilaton-driven evolution. As we stressed in Section II, to assume that
the dilaton driven phase was {\em not} completely isotropic might be
natural from the point of view of the ``asymptotic past triviality'' of
the pre-big-bang scenario \cite{5}. More specifically, the comparative
study of (spherically symmetric) gravitational collapse and
pre-big-bang scenario seem to imply the occurrence of anisotropic
dilaton driven phases. As we stressed, the example given in Section III
can be interpreted, from the point of view of the background
evolution in the Einstein frame, as the gravitational collapse of a
sphere filled with a perfect (stiff) fluid \cite{11} leading,
ultimately, to an anisotropic four-dimensional dilaton-driven phase in
the string frame picture.

The fact that anisotropic initial conditions are generally allowed by
the solutions of the tree-level action might not be, in principle,
sufficient in order to justify our assumptions. In fact what we need
for our considerations is a metric which is anisotropic not only
locally (i.e. for a short amount of time in the dilaton-driven phase)
but globally (i.e. for all the duration of the dilaton-driven
phase). What might happen is that the addition of higher string
tension corrections to the tree-level action forbid the presence of a
(long) anisotropic phase. If the anisotropic phase is not
sufficiently long the axion fluctuations will not have enough time to
fully develop their growing modes.

A fair measure of the anisotropy of a four-dimensional metric can be
given, in our particular case as 
\begin{equation}
 A(t) \equiv \frac{H-F}{\dot{V}}= \frac{3(H - F)}{H + 2 F},~~~V(t)= 
\frac{\log{[\sqrt{- g}~]}}{3},
\label{def}
\end{equation}
(this type of measure for the anisotropy of a four-dimensional
geometry was introduced long ago by Zeldovich \cite{zel}).
Notice that, in the isotropic limit (i.e. $H\rightarrow F$), 
$\dot{V}(t)$ coincides with the Hubble factor and $A(t)
\rightarrow 0$. Another possible measure of the degree of anisotropy
of a (homogeneous) background geometry can be also given as  the ratio
between the Weyl and Riemann invariants
\begin{equation}
B(t) =\frac{C_{\mu\nu\alpha\beta}C^{\mu\nu\alpha\beta}}
{R_{\mu\nu\alpha\beta}R^{\mu\nu\alpha\beta}},
\label{def2}
\end{equation}
which is supposed to vanish in the isotropic limit (see Appendix B for
the explicit expression of $B(t)$ in our background geometry).

Consider now a dilaton driven-phase evolving 
according to the solutions discussed in the previous section, i.e. 
$H\sim -7/(9 t)$ and $F\sim -4/(9t)$. Then $A(t) \sim 3/5$.
The question is then the following: if we introduce quadratic
corrections to the tree-level action will $A(t)$ be roughly constant for
the whole duration of the dilaton driven phase? In other words: can we
show that $A$ is roughly frozen to the value given by the
dilaton-driven ``vacuum''solutions prior to the onset of the stringy
phase where the curvature is stabilized? If this is not the case,
namely if $A(t)$ is not constant during the dilaton-driven phase,
then the anisotropy in the background will only be local.

Let us start from the string effective action supplemented by the first
string tension correction in four (anisotropic) dimensions. We 
follow the notation of
 \cite{14} where this problem 
was studied in the string frame picture 
(see also \cite{15} for the discussion of 
similar problems in the Einstein frame picture):
\begin{equation}
S= - \frac{1}{2 \lambda^2_{s}}\int d^4x \sqrt{- g} e^{- \phi} \biggl[ R
+ g^{\alpha\beta} \partial_{\alpha}\phi \partial_{\beta} \phi - \frac{
w\lambda_{s}^2}{4}\biggl(R_{GB}^2 - (g^{\alpha\beta} 
\partial_{\alpha}\phi\partial_{\beta}\phi)^2\biggr)\biggr],
\label{sec}
\end{equation}
where $R_{GB}^2$ is simply the Gauss-Bonnet invariant expressed in
terms of the Riemann, Ricci and scalar curvature invariants
\begin{equation}
R_{GB}^2 = R_{\mu\nu\alpha\beta}R^{\mu\nu\alpha\beta} - 4
R_{\mu\nu}R^{\mu\nu} +R^2,
\end{equation}
and $w$ is a
numerical constant of order 1 which can be precisely computed
depending upon the specific theory we deal with (for instance 
$w =-1/8$ for heterotic strings). Notice that in \cite{13} this problem
was discussed in greater generality. Here we will recall and extend
the topics which are directly relevant in our investigation.

It is convenient, for the problem at hand, not to synchronize
immediately the clocks. Therefore the metric will be
\begin{equation}
g_{\mu\nu} = {\rm diag}[ N(t)^2, -a(t)^2, -b(t)^2, -b(t)^2].
\label{met}
\end{equation}
Notice that by fixing the lapse function to $1$ corresponds to the
synchronous coordinate frame.
Inserting Eq. (\ref{met}) into Eq. (\ref{sec}) we get,after
integrating by parts the derivatives of the lapse function,
\begin{equation}
S= \frac{1}{2 \lambda_{s}^2} \int d t e^{\alpha + 2 \beta
-\phi}\biggl\{ \frac{1}{N} \biggl[ - \dot{\phi}^2 - 2 \dot{\beta}^2 -
4 \dot{\alpha} \dot{\beta} + 2 \dot{\alpha} \dot{\phi} + 4 \dot{\beta}
\dot{\phi}\biggr] + \frac{w\lambda_{s}^2}{4 N^3}\biggl[ 8 \dot{\phi}~
\dot{\alpha}~ \dot{\beta}^2 - \dot{\phi}^4\biggr]\biggr\}
\label{sec2}
\end{equation}
where we used the notation $a(t) = e^{\alpha(t)}$, $b(t) = e^{\beta(t)}$.
We can take the variation of this action with respect to $N$,
$\alpha$, $\beta$ and $\phi$ and we will get three dynamical equations
supplemented by a constraint connecting the first derivatives of the
fields. These equations are reported in Appendix B.
Here we give the final expression of the system of nonlinear
differential equations obeyed by $x \equiv \dot{\alpha}$, $y\equiv
\dot{\beta}$, $z\equiv\dot{\phi}$:
\begin{eqnarray}
\dot{x} &=& \biggl\{{\frac{-\left[ -8\,{y^2} + 8\,y\,z - 4\,{z^2} + {z^4} + 
       8\,{x^2}\,\left( -1 + y\,z \right)  + 
       8\,x\,\left( z + {y^2}\,z - 
          y\,\left( 1 + {z^2} \right)  \right)  \right] }{8\,
     \left( -1 + y\,z \right) }}\biggr\}
\nonumber\\
&+&\biggl[{\frac{1 - x\,z}{-1 + y\,z}}\biggr] \dot{y}
       +\biggl[{\frac{-1 - x\,y}{-1 + y\,z}}\biggr]\dot{z}
\label{I}\\
\dot{y} &=&\biggr[\frac{-8\,\left( 3\,{y^2} + {x^2}\,\left( 1 + {y^2} \right) 
 + 
     2\,x\,\left( y + {y^3} \right)  \right)  + 
  8\,\left( x + 2\,y \right) \,z - 4\,{z^2} + 
  4\,\left( x + 2\,y \right) \,{z^3} - 3\,{z^4}}{16(1 + x y)}\biggr]
\nonumber\\
&+&
     \biggl[{\frac{2 + 3\,{z^2}}{4 + 4\,x\,y}}\biggr] \dot{z} + 
\biggl[{\frac{-\left( 1 + {y^2} \right) }{2\,\left( 1 + x\,y \right) }}
\biggr]\dot{x}
\label{II}\\
\dot{z} &=& \biggl\{{\frac{-\left[ 16\,y\,z + 16\,{y^3}\,z 
+{z^2}\,\left( -4 + {z^2} \right)  
- 8\,{y^2}\,\left( 3 + {z^2} \right)  \right] }{8\,\left( 1 + {y^2}
\right) }}\biggr\} + 
\biggl[{\frac{2\,\left( 1 - y\,z \right) }{1 + {y^2}}}\biggr]\dot{y}
\label{III}
\end{eqnarray}
It is  known from previous studies \cite{13} 
that this system of nonlinear differential equations
admits an isotropic fixed point which can be obtained by setting
$x=y={\rm constant}$ in the constraint given in Eq. (\ref{constr})
which will give a consistency condition which leading to $x=y=
0.616...$ and $z= 1.414...$ . It is also known that this (isotropic)
fixed point can attract (mildly) anisotropic dilaton-driven
cosmologies \cite{13}. The question we
have to address, in our context, is threefold. First of all we have to ask
if the anisotropic models we are interested in from the previous
sections are attracted towards this isotropic fixed point. If this is
the case we have to investigate how the anisotropic phase is
analytically connected with the isotropic phase with constant (and
regular) curvature. Finally we have to understand quantitatively how
long is the duration of the anisotropic phase before the
isotropization (triggered by the string tension corrections) takes
place.

\begin{figure}
\begin{center}
\begin{tabular}{|c|c|}
      \hline
      \hbox{\epsfxsize = 6.5 cm  \epsffile{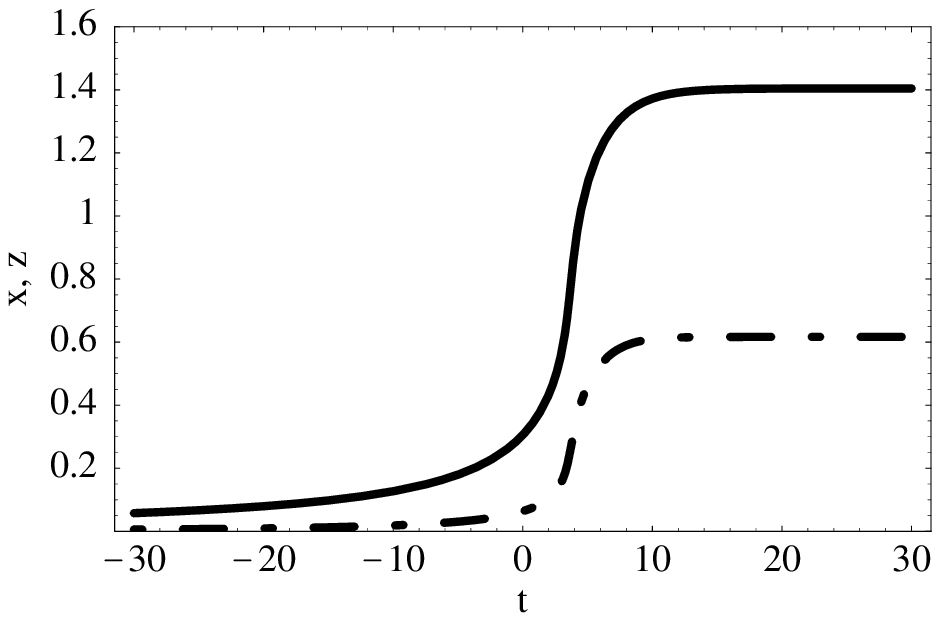}} &
      \hbox{\epsfxsize = 6.5 cm  \epsffile{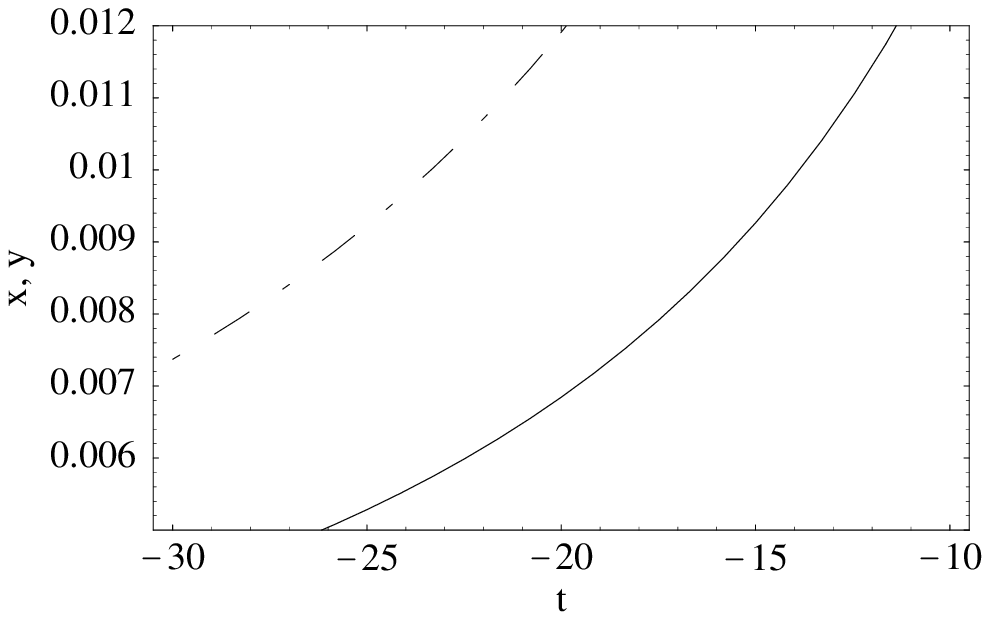}} \\
      \hline
      \hbox{\epsfxsize = 6.5 cm  \epsffile{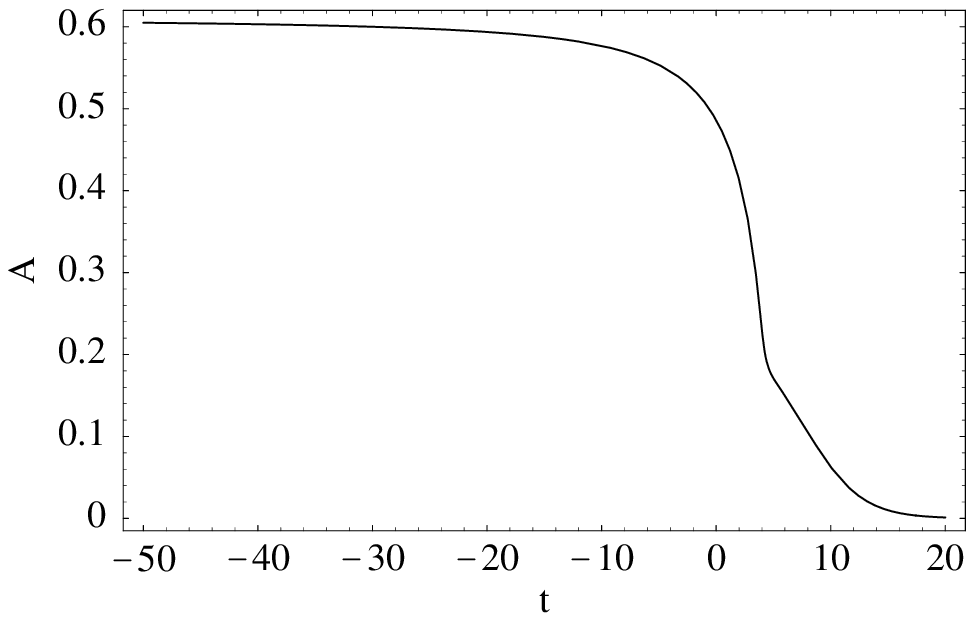}}  &
      \hbox{\epsfxsize = 6.5 cm  \epsffile{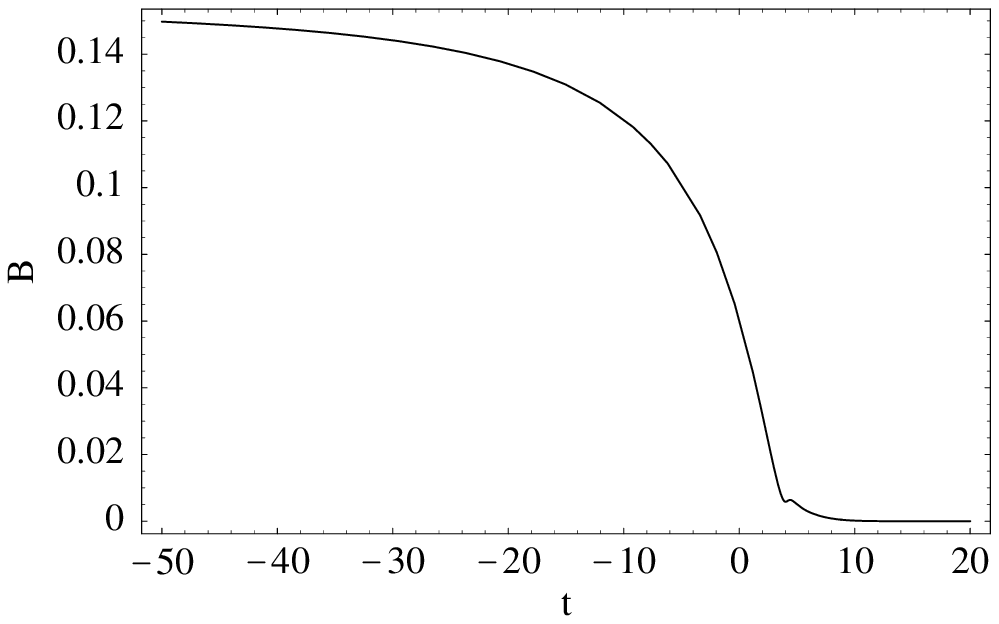}}\\
      \hline
\end{tabular}
\end{center}
\caption[a]{In the two upper plots we report the behavior of $x$, $y$
and $z$. At the top left the evolution of $z$ (thick full line) is
compared with the evolution of $x$ (dot-dashed thick line). We see that
they both freeze to a constant value given by the isotropic fixed
point. Also $y$ reaches, for large $t$ its isotropic limit but, during
the dilaton-driven phase the background is anisotropic. In the top
right plot we show indeed the evolution of $x$  (thin dot-dashed line) 
and $y$ (thin full line) prior to the
isotropization occurring at the onset of the stringy phase. This point
is also stressed in the two lower plots. In the bottom left plot we
report the numerical evaluation of the asymmetry given in
Eq. (\ref{def}). We can see that it goes to zero for large positive
times but it is roughly constant during the dilaton-driven phase. In
the bottom right plot we illustrate the same qualitative evolution
 of the ratio between
the Weyl and the Riemann invariants reported in Eqs. (\ref{def2})
and (\ref{def3}).}
\label{f5}
\end{figure}

In our present notation we have, from Eq. (\ref{def}) that 
\begin{equation}
A(t) = \frac{3[x(t) - y(t)]}{x(t) + 2 y(t)}.
\end{equation}
In order to answer the previous questions we will integrate
numerically the system given in Eqs. (\ref{I})--(\ref{III}) imposing,
at the initial integration time, $A(t_{0}) = 3/5$
which is equivalent to impose  initial conditions will 
 along the lines $\dot{\overline{\phi}}= (9/7) H$ and 
$\dot{\overline{\phi}}= (9/4) F$ in the $(\dot{\overline{\phi}},H)$
and $(\dot{\overline{\phi}},F)$ planes.
Our results are reported in Fig \ref{f5}. We see, in this particular
case, that the model, initially anisotropic, is indeed attracted
towards its isotropic fixed point by virtue of the string tension
corrections. 
The curvature invariants (including
the Weyl invariant) are all regular. Moreover the Weyl invariant and
the asymmetry defined in Eq. (\ref{def}) 
both vanish for large times signaling that the isotropy is indeed recovered.
The anisotropic and the isotropic phase are then connected and the
anisotropic regime lasts as long as $B(t)$ and $A(t)$ are almost
frozen to their constant values (fixed by the initial
conditions). This evolution is typical,
in our examples, for negative times. In this regime the solutions are
the ones discussed in Section II and, therefore the typical behavior
of the axionic growing modes is different from the one dictated by a
completely isotropic dilaton driven dynamics and leading to red
logarithmic energy spectra. By solving
Eqs. (\ref{I})--(\ref{III}) together with Eq. (\ref{psieq}), we
checked that, as far as the dilaton driven phase is concerned, the
axionic growing modes are indeed the ones we computed by using the
tree-level solutions discussed in Section II. We will report this study
elsewhere. 

One can ask, at this point, how general is this behavior. Our answer
is that it holds for various models which we would call mildly
anisotropic, from the point of view of their initial (primordial)
asymmetry. The results we just presented in a particular case do not
hold in the case of strongly anisotropic initial conditions where, for
example, one of the two scale factors contracts (in the string frame)
instead of expanding like the ones appearing in the second quadrant of
the plot reported in Fig. \ref{f1}, like, for instance the case where
$\alpha = -1/3$ and $\beta = 2/3$. Since for the  (anisotropic)
dilaton-driven solutions discussed in Section II we have that the
asymmetry is 
\begin{equation}
A(t_0)=3 \frac{(\alpha - \beta)}{\alpha + 2 \beta} 
\label{as}
\end{equation}
we have for $\alpha = -1/3$ and $\beta = 2/3$ that $A(t_0)= -3$. With
these initial conditions the system of Eqs. (\ref{I})--(\ref{III})
evolves towards a singularity, the isotropic fixed point is not
reached and the anisotropic phase lasts forever. Moreover, we can see
that the asymmetry grows (im modulus) becoming more and more
negative. In this case the string tension corrections will make the
geometry more and more anisotropic.

\begin{figure}
\begin{center}
\begin{tabular}{|c|c|}
      \hline
      \hbox{\epsfxsize = 6.5 cm  \epsffile{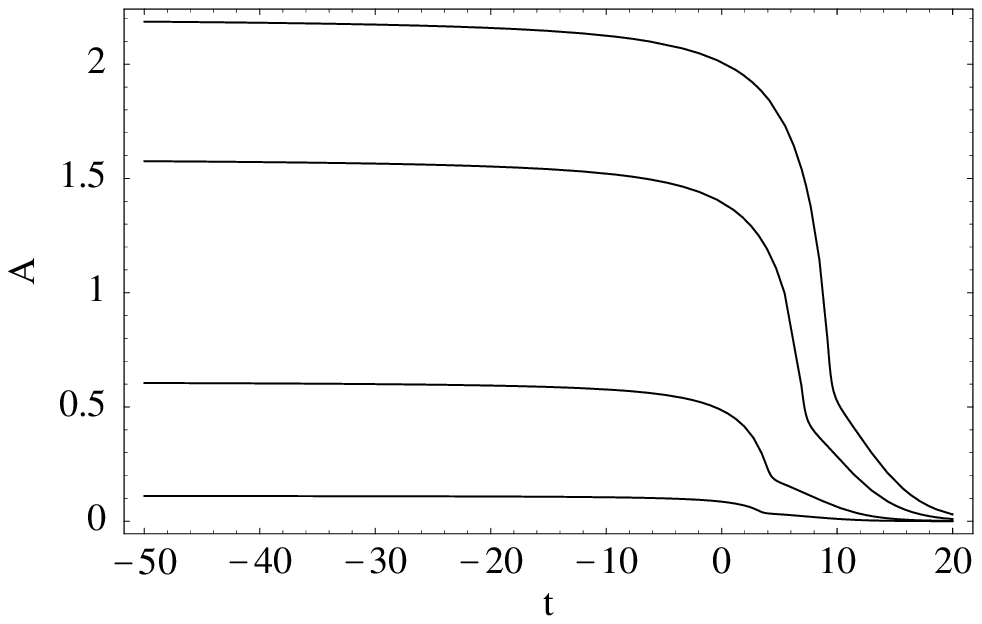}} &
      \hbox{\epsfxsize = 6.5 cm  \epsffile{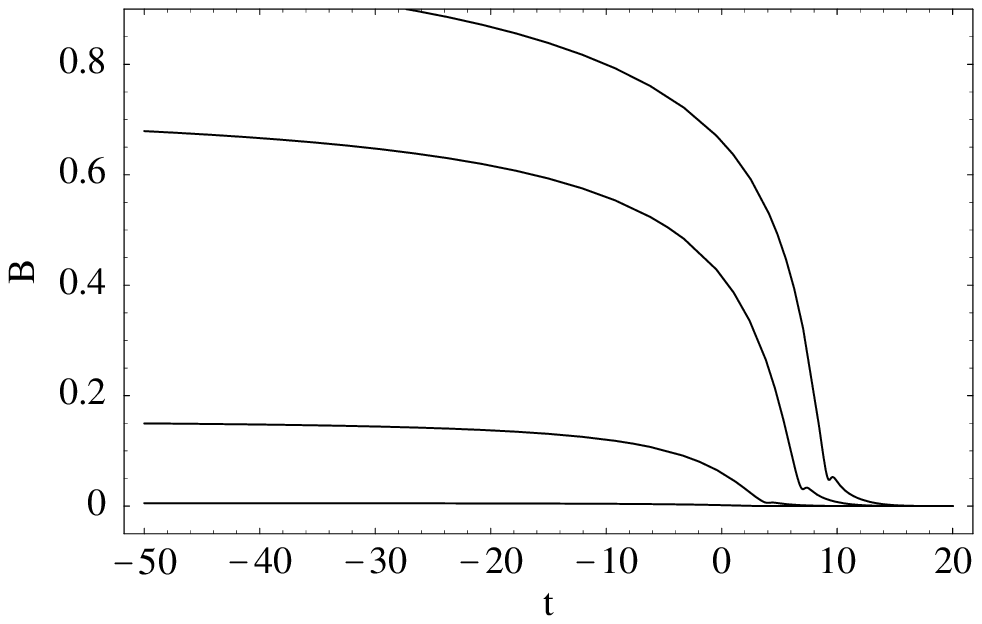}} \\
      \hline
\end{tabular}
\end{center}
\caption[a]{Behavior of the asymmetries $A(t)$ and $B(t)$ for
different initial conditions within the region of the $(\alpha,\beta)$
plane leading to blue axion spectra.}
\label{f6}
\end{figure}

Therefore we  studied the accelerated branch of the dilaton driven
solutions (i.e. $\alpha<0$ and $\beta<0$) and we fixed the  initial
conditions of  the system  (\ref{I})--(\ref{III}) by requiring
that 
\begin{equation}
A(t_0) = 3 \frac{(-\sqrt{1 - 2 \beta^2} - \beta)}{-\sqrt{1 - 2
\beta^2} + 2 \beta}
\label{as2}
\end{equation}
(where we simply inserted $\alpha = - \sqrt{1 - 2 \beta^2} $ 
into Eq. (\ref{as}) in order
to force the initial conditions to lie inside the space of the
solutions of the tree level action [see also Fig. \ref{f1}]).
We then moved $\beta$ in the range $-0.333(3)\laq \beta<0$ where,
according to Fig. \ref{f1}, blue
axionic spectra are expected on the basis of the tree-level dilaton
driven  solutions. Our results are reported in Fig. \ref{f6} in terms
of the two previously introduced estimators of the anisotropy $A(t)$
and $B(t)$. We plot four cases namely $\beta= -0.333(3), -0.222(2),
-0.11(1), -0.55(5)$ and we find the same behavior discussed in
Fig. \ref{f5} though with different initial asymmetry. Notice that the model
with $\beta = -0.555(5)$ does not give blue spectra but, nonetheless,
it is driven towards isotropy for large positive times. 
 We recall that the motivation of
this last section was to understand if string tension corrections can
prevent the existence of anisotropic phases. In this sense, our
conclusion is that string tension corrections do not prevent the
existence of long anisotropic dilaton-driven epochs. Moreover, we
showed that the models leading to blue axion spectra are likely to be
attracted towards isotropic fixed points as a result of the string
tension corrections.

\renewcommand{\theequation}{6.\arabic{equation}}
\setcounter{equation}{0}
\section{Concluding Remarks and Speculations}

\begin{figure}
\centerline{\epsfxsize = 7 cm  \epsffile{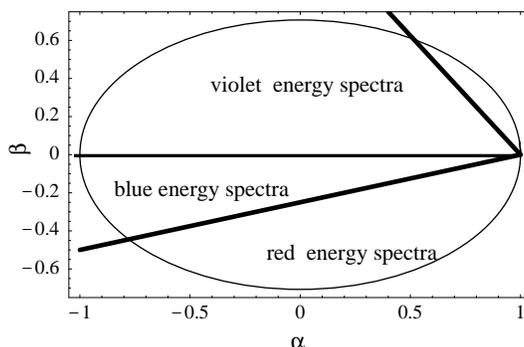}} 
\caption[a]{Axionic logarithmic energy spectra for different
dilaton-driven models with Kasner-like exponents $(\alpha,\beta)$ in
the string frame.}
\label{f7}
\end{figure}

In this paper we relaxed the isotropy assumption in the study of the
amplification of the axionic fluctuations in string cosmology. In the
isotropic case the four-dimensional spectra are typically red. 
We then focused our attention on the fully anisotropic
(four-dimensional) dilaton driven solutions where 
 the growing modes of the axion fluctuations
are completely different from the one occurring in the completely
isotropic case with the same number of dimensions. Assuming (as in the
isotropic case) a smooth transition to a radiation dominated phase,
the obtained axion
spectra have an amplitude which depends upon the primordial anisotropy
of the space-time encoded in the asymmetry between longitudinal and
transverse momenta which can be expressed in terms of a primordial 
rapidity. The spectral slope is determined by the growing mode
solution and can be either flat or blue for anisotropic models whose 
scale factors are all expanding though at a different rate. Our
conclusions are summarized in Fig. \ref{f7}.
As we showed the situation is qualitatively different if compared with
the  isotropic dilaton-driven evolution where only red spectra
are possible.  We focused our attention on blue spectra because they
are produced by models with expanding four-volume and growing dilaton
coupling. Red and violet spectra seem less natural from the
anisotropic point of view since they require that at least one of the
scale factors contracts (in the string frame). We compared our
phenomenological point of view with anisotropic models occurring as a
result of spherically symmetric gravitational collapse and we showed
that some of those models lead to blue axionic energy spectra. 
This conclusion extends and complements previous works on the subject 
\cite{4b,4c}.

Motivated by our phenomenological considerations we investigated the
influence of string tension correction on the evolution of anisotropic
dilaton-driven epochs. We found that string tension corrections never
prevent the existence of anisotropic phases. Moreover if the
anisotropy is mild (like in the region of Fig. \ref{f7} compatible
with blue spectra) all the pre-big bang phase is anisotropic but it is
attracted for large positive times towards an isotropic fixed
point. 

What we presented in this paper seem to suggest that it might be
plausible to consider more seriously the possible effect of primordial
anisotropies  on the axionic spectra as it was recently done in the 
purely isotropic case \cite{4b}. It is not impossible to imagine that
the amplitude of our spectra (and then the primordial ratio between
longitudinal and transverse momenta) can be constrained by the temperature
fluctuations in the Cosmic Microwave Background radiation.
More generally, we find interesting that our
considerations seem to point towards a more detailed analysis of 
anisotropic string cosmological models.

\section*{Acknowledgments}

I would like to thank G. Veneziano for very useful discussions.

\newpage 

\begin{appendix}

\renewcommand{\theequation}{A.\arabic{equation}}
\setcounter{equation}{0}
\section{Mixing coefficients}

In this appendix we report the mixing coefficients employed in the
calculation of the energy spectrum of the amplified axionic
perturbations discussed in Section III. Defining the function 
\begin{equation}
S(r) = \frac{1}{ \sqrt{ r^2 + 1}}, ~~~r=
\frac{k_{T}}{k_{L}} \equiv\frac{\omega_{T}}{\omega_{L}}
\end{equation}
we have that the mixing coefficients are

\begin{eqnarray}
c_{-}(k) &=& \frac{[S(r)]^{5/2}\sqrt{x} }{ 2 \sqrt{\pi}} 
e^{-i x [ 2 + S(r)]} \{ (- 3 - 4 i x + \frac{2 i x}{ S(r)} - r^2 ( 3 +
4i x)) U[ \frac{3}{2} - \frac{i}{2} \epsilon r, 3, \frac{2 i x}{ S(r)}] 
\nonumber\\
&+& \frac{2 x}{S(r)} (3 i + \epsilon r)  
U[ \frac{5}{2} - \frac{i}{2} \epsilon r, 4 ,
2 i x S(r)]\}
\nonumber\\
c_{+}(k) &=& \frac{ [S(r)]^{5/2}\sqrt{x}}{ 2 \sqrt{\pi}}
e^{-i x [ 2 - S(r)]} \{ (- 3 + 4 i x + \frac{2 i x}{ S(r)} + r^2 ( -3 + 4
i x ))U[ \frac{3}{2} -\frac{i}{2} \epsilon r , 4, 2 i x  S(r) ]
\nonumber\\
&+& \frac{2 x}{S(r)} (  3 i + \epsilon r) U[ \frac{5}{2} - \frac{i}{2}
\epsilon r, 4, 2 i x S(r)]\}
\end{eqnarray}
where $x= k\eta$.

\renewcommand{\theequation}{B.\arabic{equation}}
\setcounter{equation}{0}
\section{Equations of motion with string tension corrections in the
fully anisotropic case}
Taking the variation of the actin given in Eq. (\ref{sec2}) with
respect to the lapse function $N(t)$ and imposing, afterwards, the
cosmic time gauge we get the constraint
\begin{equation}
\dot{\phi}^2 + 2 \dot{\beta}^2 + 4 \dot{\alpha}\dot{\beta} 
- 2 \dot{\alpha}\dot{\phi} - 4 \dot{\beta} \dot{\phi} +
\biggl[\frac{3}{4} \dot{\phi}^4 - 6 \dot{\alpha} \dot{\phi} 
\dot{\beta}^2\biggr]=0
\label{constr}
\end{equation}
where we took string units $\lambda_s=1$ and $w=1$.
By varying the action with respect to $\alpha$, $\beta$ and $\phi$ we
get, for $N(t)=1$,
\begin{eqnarray}
&& 4 \ddot{\beta} - 2 \ddot{\phi} + (4 \dot{\beta} - 2 \dot{\phi})
(\dot{\alpha} + 2 \dot{\beta} - \dot{\phi}) - 2 \biggl[ (\dot{\alpha} +
2 \dot{\beta} - \dot{\phi})\dot{\phi} \dot{\beta}^2 
+ \ddot{\phi} \dot{\beta}^2+ 2
\dot{\beta} \ddot{\beta}\dot{\phi}\biggr] +
L(t)=0,
\\
&& 4 (\ddot{\beta} + \ddot{\alpha} - \ddot{\phi}) + 4( \dot{\alpha} +
\dot{\beta} - \dot{\phi})( \dot{\alpha} + 2 \dot{\beta} -\dot{\phi}) -
4 \biggl[ \ddot{\beta}\dot{\alpha} \dot{\phi} + \dot{\beta}
\ddot{\alpha} \dot{\phi} + \dot{\beta} \dot{\alpha} \ddot{\phi} + 
\dot{\beta} \dot{\alpha} \dot{\phi} (\dot{\alpha} + 2 \dot{\beta} -
\dot{\phi})\biggr] + 2 L(t)=0,
\\
&&
2 (\ddot{\phi} - \ddot{\alpha} - 2 \ddot{\beta}) + 2 (\dot{\phi} -
\dot{\alpha} -2 \dot{\beta})(\dot{\alpha} + 2 \dot{\beta} -
\dot{\phi}) 
\nonumber\\
&&-\biggl[   2 \dot{\beta}(\ddot{\alpha}\dot{\beta} + 2 \dot{\alpha}
\ddot{\beta}) + (2 \dot{\alpha} \dot{\beta}^2 -
\dot{\phi}^3)(\dot{\alpha} + 2 \dot{\beta} - \dot{\phi}) - 3
\dot{\phi}^2 \ddot{\phi} \biggr] -L(t)
=0,
\end{eqnarray}
where we defined 
\begin{equation}
L(t) = -\dot{\phi}^2 -2 \dot{\beta}^2 - 4 \dot{\alpha}\dot{\beta} + 2
\dot{\alpha} \dot{\phi} + 4 \dot{\beta}\dot{\phi} + \frac{1}{4}(8
\dot{\phi}\dot{\alpha} \dot{\beta}^2 - \dot{\phi}^4).
\end{equation}
Defining now 
\begin{equation}
x(t) = \dot{\alpha},~~~~~~y(t) = \dot{\beta}, ~~~~~~z(t)= \dot{\phi}
\end{equation}
we get, after trivial algebra, the equations reported in Section VI.

In Section VI we also used an alternative measure of the anisotropy of
a homogeneous space-time, namely the ratio between the Weyl and
Riemann invariants. In the specific case of our metric this quantity
reads:
\begin{equation}
B(t) =\frac{C_{\mu\nu\alpha\beta}C^{\mu\nu\alpha\beta}}
{R_{\mu\nu\alpha\beta}R^{\mu\nu\alpha\beta}} \equiv \frac{1}{3}\frac{[
x y -x^2 -\dot{x} + \dot{y}]^2}{[x^4 + 2 x^2 y^2  + 3 y^4 + 2 x^2
\dot{x} + \dot{x}^2 + 4 y^2 \dot{y} + 2 \dot{y}^2]}
\label{def3}
\end{equation}
We can clearly see from the above expression that in the isotropic
limit (i.e. $x= y$) $B(t) \rightarrow 0$, as expected.

\end{appendix}

\end{document}